\documentclass[twocolumn,pra,aps,superscriptaddress,english,floatfix]{revtex4-1}
\usepackage{amsmath}
\usepackage{amssymb}
\usepackage{amsfonts}
\usepackage[pdftex]{graphicx}
\usepackage[dvipsnames]{xcolor}
\usepackage{subfigure} 
\usepackage[english]{babel}
\usepackage{braket}
\usepackage{color}
\usepackage{mathtools}
\usepackage{threeparttable}
\usepackage{ragged2e}
\emergencystretch=\maxdimen
\usepackage[colorlinks,linkcolor=blue,anchorcolor=blue,citecolor=blue,urlcolor=blue]{hyperref}
\usepackage{babel}
\makeatother

\begin{document}

\preprint{APS/123-QED}

\title{ Non-Dissipative Non-Hermitian Dynamics and Exceptional Points in Coupled Optical Parametric Oscillators}

\author{Arkadev Roy}
\affiliation{%
Department of Electrical Engineering, California Institute of Technology, Pasadena, California 91125, USA}%
\author{Saman Jahani}%
\affiliation{%
Department of Electrical Engineering, California Institute of Technology, Pasadena, California 91125, USA}%
 \author{Qiushi Guo}
 \affiliation{%
Department of Electrical Engineering, California Institute of Technology, Pasadena, California 91125, USA}%
 \author{Avik Dutt}
 \affiliation{%
E. L. Ginzton Laboratory, Stanford University, Stanford, California 94305, USA}%
 \author{Shanhui Fan}
 \affiliation{%
E. L. Ginzton Laboratory, Stanford University, Stanford, California 94305, USA}%
 \author{Mohammad-Ali Miri}
 \affiliation{%
Department of Physics, Queens College, City University of New York, New York, New York 11367, USA}%
 \author{Alireza Marandi}
 \email{marandi@caltech.edu}
\affiliation{%
Department of Electrical Engineering, California Institute of Technology, Pasadena, California 91125, USA}%

%

\begin{abstract}
Engineered non-Hermitian systems featuring exceptional points can lead to a host of extraordinary phenomena in diverse fields ranging from photonics, acoustics, opto-mechanics, electronics, to atomic physics. Here we introduce and present non-Hermitian dynamics of coupled optical parametric oscillators (OPOs) arising from phase-sensitive amplification and de-amplification, and show their distinct advantages over conventional non-Hermitian systems relying on laser gain and loss. OPO-based non-Hermitian systems can benefit from the instantaneous nature of the parametric gain, noiseless phase-sensitive amplification, and rich quantum and classical nonlinear dynamics. We show that two coupled OPOs can exhibit spectral anti-PT symmetry and an exceptional point between its degenerate and non-degenerate operation regimes. To demonstrate the distinct potentials of the coupled OPO system compared to conventional non-Hermitian systems, we present higher-order exceptional points with two OPOs, tunable Floquet exceptional points in a reconfigurable dynamic non-Hermitian system, and generation of squeezed vacuum around exceptional points, all of which are not easy to realize in other non-Hermitian platforms. Our results show that coupled OPOs are an outstanding non-Hermitian setting with unprecedented opportunities in realizing nonlinear dynamical systems for enhanced sensing and quantum information processing.

\end{abstract}

\maketitle
Non-Hermitian systems with engineered gain and dissipation have attracted a lot of attention thanks to their remarkable properties and functionalities which are absent in their counterparts based on closed Hermitian setups \cite{miri2019exceptional, ozdemir2019parity}. A plethora of phenomena have spawned out by judiciously manipulating these non-Hermitian physical systems namely, spontaneous parity-time symmetry breaking \cite{miri2019exceptional}, unidirectional invisibility \cite{lin2011unidirectional}, coherent perfect absorption and lasing \cite{longhi2010pt,wong2016lasing}, single mode lasing \cite{feng2014single}, generation of structured light with controllable topological charge of the orbital angular momentum mode \cite{miao2016orbital}, to name a few. \

Non-Hermitian systems are often characterized by the presence of an exceptional point (EP), where the eigenvalues and eigenvectors simultaneously coalesce (non-Hermitian degeneracies), and have been explored in the context of parity-time symmetric systems with balanced gain/loss and even in purely dissipative arrangements. The presence of an EP lead to several counter-intuitive phenomenon including loss induced lasing \cite{peng2014loss, el2014exceptional}, breakdown of adiabaticity \cite{hassan2017dynamically,doppler2016dynamically}, lasing without inversion \cite{doronin2019lasing}. However, most non-Hermitian optical systems realize gain/dissipation by deploying laser gain which limits its viability in certain spectral regions \cite{peng2014parity}. \\

Here we utilize parametric amplification and de-amplification in coupled OPOs to implement EP in parametric non-Hermitian systems \cite{antonosyan2015parity, el2015optical, wang2019non}, thereby presenting a system that can exhibit unique phenomena not observed in their laser-gain based counterparts. Parametric non-Hermitian systems can extend beyond the spectral coverage of laser gain \cite{miri2016nonlinearity}, and  the instantaneous nature of parametric gain also enables the realization of tunable/reconfigurable non-Hermitian systems that are otherwise difficult to achieve in conventional optics based non-Hermitian setups. We leverage this tunable aspect of parametric gain to realize interesting functionalities, which has largely been ignored in previous studies pertaining to parametrically driven non-Hermitian systems \cite{antonosyan2015parity, el2015optical, wang2019non}. Fundamentally, the presented OPO-based non-Hermitian system is in sharp contrast with conventional optical systems and can enable unique opportunities for sensing, non-Hermitian nonlinear dynamics, and quantum information processing.
\\

EPs in non-Hermitian systems have been extensively studied for potentially enhanced sensing capabilities \cite{hodaei2017enhanced, lai2019observation, chen2017exceptional}. Inspite of the underlying high sensitivity near an EP, these class of sensors relying on the resonant frequency splitting are not capable of improving the SNR (Signal-to-Noise Ratio) owing to the non-orthogonality of eigenvectors near an EP \cite{lau2018fundamental,langbein2018no,Chen_2019}. This leads to Peterman factor limited sensing \cite{wang2019petermann} where the noise is enhanced proportional to the signal enhancement, thereby limiting the efficacy of these class of sensors for quantum limited sensing \cite{degen2017quantum}. Fluctuations accompanying the gain/dissipation in conventional non-Hermitian systems limits the achievable precision. In fact, it has been shown that any linear reciprocal sensor is bounded in terms of SNR performance, and the conventional EP based sensing cannot surpass this limit \cite{lau2018fundamental}. Recently, a sensing protocol that does not measure the eigen-frequency splitting but rather measures the superposition of output quadratures using heterodyne detection, has shown the possibility of alleviating the problem of noise enhancement and realize EP enhanced sensing when operated near the lasing threshold \cite{zhang2019quantum}. The noiseless nature of phase-sensitive degenerate parametric amplification motivates studying non-Hermitian dynamics of coupled OPOs for sensing. In this regard, we explore the possibility of reduced uncertainty of fluctuations manifested in the form of squeezed noise in the vicinity of parametric EP to leverage the high sensitivity of EP in the pursuit of obtaining high SNR. It must be noted that phase sensitive parametric gain based systems are not bounded by the limit outlined in Ref. \cite{lau2018fundamental}. \\

Non-Hermitian dynamics of coupled OPOs can be extended to the nonlinear regime which can lead to several intriguing possibilities. It has been previously shown that the interplay of nonlinearity and gain/loss in conventional non-Hermitian systems can result in unidirectional transport \cite{ramezani2010unidirectional}, one parameter family of solitons \cite{abdullaev2011solitons} (in contrast to an isolated attractor based dissipative solitons) in parity-time symmetric systems, robust wireless power transfer \cite{assawaworrarit2017robust}. Previous studies implementing parametric amplification to realize non-Hermitian systems have focused on the linear dynamics only \cite{antonosyan2015parity, el2015optical, wang2019non}. We exploit rich nonlinear dynamics in our parametric non-Hermitian system (operating in the parametric oscillator regime) arising from the interplay of phase sensitive gain and the gain saturation owing to the signal to pump back-conversion. \\

The presented coupled OPO system is also an appealing platform to investigate quantum non-Hermitian physics. Previous studies of the quantum behaviour in non-Hermitian systems have identified the criticality of information flow between system and environment around the EP in a parity-time symmetric system \cite{kawabata2017information}, shift of the position of Hong-Ou-Mandel dip \cite{klauck2019observation}, and delaying of entanglement sudden death near an EP \cite{PhysRevA.100.063846}. We demonstrate non-classical behaviour including quadrature squeezing and tunable squeezing of parametric EP. These behaviors may also be extended to the non-Gaussian regime\cite{onodera2018nonlinear}. \\

\section{Model of Coupled OPOs}
We consider a system of evanescently coupled degenerate OPOs as illustrated in Fig. 1a.  The coupled-mode equations governing our system is given by:
\begin{subequations}
\begin{equation}
    \frac{da}{dt} =-\gamma_{1}a + i\Delta_{1}a +ga^{*} -g_{s_{1}}|a|^{2}a+i\kappa b 
  \end{equation} 
 \begin{equation}
        \frac{db}{dt} =-\gamma_{2}b + i\Delta_{2}b +fe^{i\phi}b^{*} -g_{s_{2}}|b|^{2}b+i\kappa a
 \end{equation} 
\end{subequations} The OPOs considered are phase matched to oscillate around the half-harmonic frequency \cite{PhysRevA.94.063809}. The continuous-wave (CW) pump is non-resonant and its dynamics is adiabatically eliminated. The signal field envelopes in the two resonators are designated by $a$ and $b$ respectively. The signal in the first resonator experiences a round-trip loss (intrinsic+out-coupling) of $\gamma_{1}$, a detuning of $\Delta_{1}$, and a parametric gain of $g$ provided by the non-resonant pump. The gain can be assumed constant for the frequency range of interest around the half-harmonic frequency. The parametric gain is phase sensitive, and the phase of the pump driving the first resonator is taken as reference. The gain saturation term is denoted by $g_{s_{1}}$ which originates from the signal to pump back-conversion due to second-harmonic generation which is the reverse of the down-conversion process. The strength of the dispersive coupling is represented by $\kappa$. Similar terms appearing in Eq.(1b) describe associated quantities in the second resonator. The pump driving the parametric interaction in the second resonator is phase shifted by $\phi$ as compared to the first pump. The parametric gain is proportional to the pump strength and is given by $f$. Both the pumps are at $2\omega_{0}$, where $\omega_{0}$ is the half harmonic frequency. The time scale is normalized to the round-trip time. 
\begin{figure}[!h]
\centering
\includegraphics[width=0.42\textwidth]{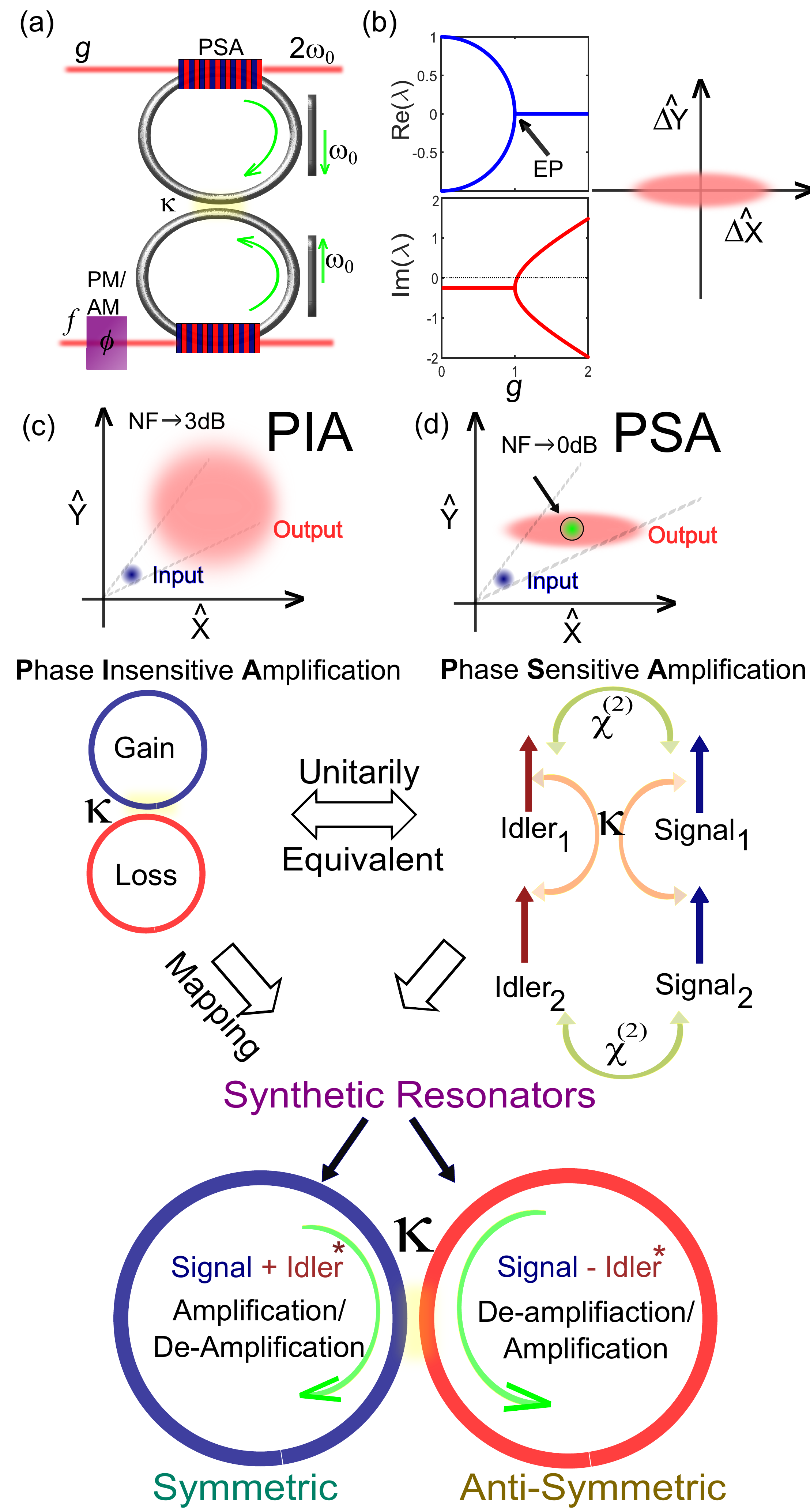}

\caption{\label{fig: schematic} \textbf{Schematic depicting the concept of parametric  EP}. a) Coupled OPO with evanascent coupling $\kappa$, driven by non-resonant pumps at $2\omega_{0}$. The strength of the drive in the first OPO is $g$ and the strength of the drive in the second OPO is $f$. The pumps can be phase (PM)/amplitude (AM) modulated, and the resonant signals at half-harmonic ($\omega_{0}$)  are extracted from the respective out-couplers. b) The appearance of an EP as the parametric gain parameter $g$ is varied at a fixed coupling rate $\kappa$. In the vicinity of this EP apart from enhanced responsivity due to the branch point singularity nature of EP, there exists squeezed noise which can reduce the uncertainty of the output field below the standard noise limit in a suitable quadrature.   c) Conventional non-Hermitian systems employs phase insensitive amplification (PIA) mechanism and thus the noise figure of the system under consideration cannot go below 3dB. d) We leverage the phase sensitive amplification (PSA) in the realization of parametric EP which can ideally approach noiseless amplification. We map the usual gain-loss coupling in conventional non-Hermitian system to phase sensitive quadrature parametric amplification and de-amplification and represent the equivalent process in synthetic resonators.}
\end{figure}

 We assume that the resonators are identical in terms of the loss ($\gamma$) and gain-saturation ($g_{s}$) terms for simplicity. This can be achieved by accessing the two degrees of freedom of a single resonator namely the clockwise and counter-clockwise propagation modes. In the absence of these assumptions the results discussed in this work will still hold true, albeit some quantitative differences. 
 
 \section{Linear Dynamical Analysis}
 There are two regimes of parametric oscillation, namely the non-degenerate regime and the degenerate regime \cite{roy2020spectral}. In the degenerate regime the system oscillates at $\omega_{0}$, while in the non-degenerate regime owing to energy conservation constraint the system oscillates with symmetric sidebands centered around $\omega_{0}$. First, we consider that both the half-harmonic signals are on resonance, i.e. $\Delta_{1}=\Delta_{2}=0$.  \\
 
 In the non-degenerate regime (under the scope of linearized analysis i.e. ignoring gain saturation) we can consider the following ansatz for the signal envelopes in the two resonators as: 
 \begin{subequations}
\begin{equation}
    a= A e^{(\lambda_{I}-i\lambda_{R})t}+B e^{(\lambda_{I}+i\lambda_{R})t}
  \end{equation} 
 \begin{equation}
        b= C e^{(\lambda_{I}-i\lambda_{R})t}+ D e^{(\lambda_{I}+i\lambda_{R})t}
 \end{equation} 
 where $A$ and $B$ represent the complex envelopes for the symmetric primary sidebands for resonator 1, and $C$ and $D$ represent the same for resonator 2. Here, the real part of eigenvalues ($\lambda_{R}$) corresponds to the spectral splitting, while the imaginary part ($\lambda_{I}$) is related to the growth/decay rate. $A, C$ can also be read as the signals and $B, D$ as the idlers. The eigenvalues can be obtained from the following equation:
 \begin{gather}
 (\lambda_{R}+i\lambda_{I})\begin{bmatrix} A \\ B^{*}  \\ C \\ D^{*}\end{bmatrix}
 =  \begin{bmatrix}
  -i\gamma & ig & -\kappa & 0 \\
 ig & -i\gamma & 0 & \kappa \\
 -\kappa & 0 & -i\gamma & ife^{i\phi} \\
 0 & \kappa & ife^{-i\phi} & -i\gamma
   \end{bmatrix} 
   \begin{bmatrix} A \\ B^{*}  \\ C \\ D^{*}\end{bmatrix}
\end{gather}
\end{subequations}
\begin{figure}[!h]
\centering
\includegraphics[width=0.45\textwidth]{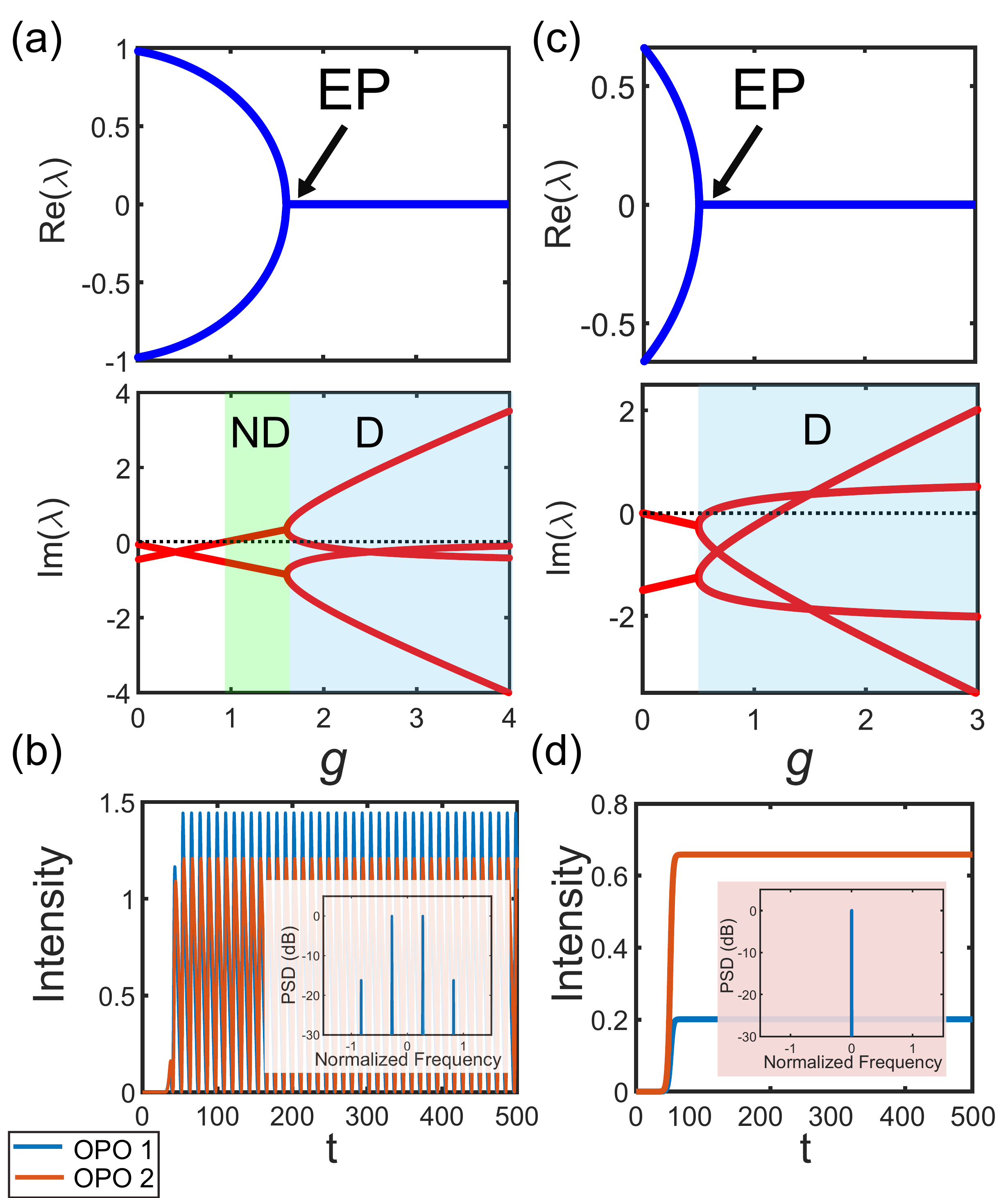}
\caption{\label{fig: schematic}\textbf{Parametric Oscillation in Coupled OPO in the presence of EP}. a,b) Coupled OPO system initiates parametric oscillation in the non-degenerate (ND) phase at threshold. The parameters used are: $f=0.4$, $\kappa=1$, $g_{s}=0.3$, and $\gamma = 0.25$. At higher values of $g$ above threshold the system operates in degenerate (D) phase. In a) the real and imaginary part of the eigenvalues ($\lambda$) are shown that emanates from the linearized analysis. The threshold is indicated by the dashed black line.  The time domain profile of the intra-cavity intensity (for $g=1.5$) in both the resonators are shown in b).  The corresponding spectral domain information appearing in the inset confirms the non-degenerate oscillation phase.   c,d) Coupled OPO system initiates parametric oscillation in the degenerate phase at threshold. The parameters used are: $f=1.5$, $\kappa=1$, and $\gamma = 0.75$. There is no occurrence of non-degenerate oscillation. In c) the real and imaginary part of the eigenvalues are shown that emanates from the linearized analysis.  The time domain profile of the intra-cavity intensity (for $g=1$) in both the resonators are shown in d).  The corresponding spectral domain information appearing in the inset confirms the degenerate oscillation phase.}
\end{figure}

The underlying Hamiltonian of the coupled OPO system exhibits spectral anti-PT symmetry \cite{antonosyan2015parity}. The Hamiltonian governs the dynamics as: $i\frac{d\Tilde{V}}{dt}= H\Tilde{V}$, where $\Tilde{V}=\left[ \Tilde{A}, \Tilde{B}^{*}, \Tilde{C},\Tilde{D}^{*} \right]^{T}$, $\Tilde{A}=A e^{(\lambda_{I}-i\lambda_{R})t}$, $\Tilde{B}=B e^{(\lambda_{I}+i\lambda_{R})t}$, $\Tilde{C}=C e^{(\lambda_{I}-i\lambda_{R})t}$, and $\Tilde{D}=D e^{(\lambda_{I}+i\lambda_{R})t}$. The discrete symmetry of the system can be expressed as: $P_{1}P_{2}TH=-HP_{1}P_{2}T,$ where $T$ is the time reversal operator, and the parity operators action in the spectral domain is defined by: $P_{1}=\{A \leftrightarrow B^{*} \}$ and $P_{2}=\{C \leftrightarrow D^{*} \}$. The system dynamics is also unitarily equivalent to a PT symmetric system, where the unitary transformation $\mathbb{U} = $ $\frac{1}{\sqrt{2}}\begin{bmatrix} 1 & -1 & 0 & 0 \\ 0 & 0 & 1 & -1 \\ 1 & 1 & 0 & 0 \\ 0 & 0 & 1 & 1 \end{bmatrix}$, such that $H_{\textrm{PT}}=\mathbb{U}H\mathbb{U}^{\dagger}$. This mapping is shown schematically in Fig. 1c and 1d. The signals of the two OPOs are coupled by the evanescent linear coupling $\kappa$, so do the idlers. While the signal and the idler within the same OPO are coupled nonlinearly by the nonlinear phase sensitive coupling engendered by $\chi^{(2)}.$ Under the said unitary transformation ($\mathbb{U}$), this process can be mapped to a PT symmetric system of coupled synthetic resonators with the positive superposition of the signal and the idler conjugated fields experiencing amplification, while the negative superposition of the signal and idler fields get de-amplified. It should be noted that due to the onset of nonlinearity arising due to back conversion ($g_{s}$) additional sidebands will appear in the complete nonlinear solution. \\
 
 In the degenerate regime the signals in both resonators are half harmonics. Here we can express the signal evolution in terms of their quadrature components. We define $X_{1}= (a+a^{*}), Y_{1}=\frac{a-a^{*}}{i}$ and $X_{2}= (b+b^{*}), Y_{2}=\frac{b-b^{*}}{i}$. These quadrature components evolve as $e^{\lambda_{I}t}$. The eigenvalues can be obtained from the following evolution equation:
\begin{widetext}
\begin{gather}
 i\lambda_{I}\begin{bmatrix} X_{1}\\ Y_{1}  \\ X_{2} \\ Y_{2}\end{bmatrix}
 =  \begin{bmatrix}
  -i\gamma + ig & 0 & 0 &-i\kappa  \\
 0 &   -i\gamma-ig & i\kappa & 0 \\
 0 & -i\kappa &  -i\gamma+if\textrm{cos}(\phi) & if\textrm{sin}{\phi} \\
  i\kappa & 0 & if\textrm{sin}{\phi} & -i\gamma -if\textrm{cos}{\phi}
   \end{bmatrix} 
   \begin{bmatrix} X_{1}\\ Y_{1}  \\ X_{2} \\ Y_{2}\end{bmatrix}
\end{gather}
\end{widetext}

The transition from the non-degenerate oscillation regime to the degenerate oscillation regime is marked by the presence of an exceptional point. This point in the parameter space is characterized by the simultaneous collapse of eigenvectors and the coalescence of the eigenvalues. The disparity between the geometric and the algebraic multiplicity at the exceptional point is determined by the order of the exceptional point. 
\section{Results}
The threshold for parametric oscillation in the coupled OPO is determined by the linear eigenvalues i.e. $\lambda_{I}= 0$, with oscillation occurring for $\lambda_{I}> 0$ \cite{sup}. This extra caution is because of the possibility of occurrence of oscillation self-termination \cite{sup} analogous to laser self-termination \cite{peng2014loss, el2014exceptional}. Just above threshold the system of coupled OPO's can oscillate either in non-degenerate (Fig. 2a and 2b) or in degenerate mode (Fig. 2c and 2d). However, far above threshold the effect of nonlinearity becomes significant and the system is no longer governed by the linearized dynamics (Eq. 2,3). In this regime nonlinearity can induce  a phase transition from non-degeneracy to degeneracy as shown in Fig.3b and 3c, similar to laser systems  \cite{PhysRevLett.111.263901, hassan2015nonlinear}. This transition resembles a soft/ super-critical bifurcation. Analytical results depicting this phenomenon for a representative case is shown in \cite{sup}. \\
\begin{figure}[!h]
\centering
\includegraphics[width=0.45\textwidth]{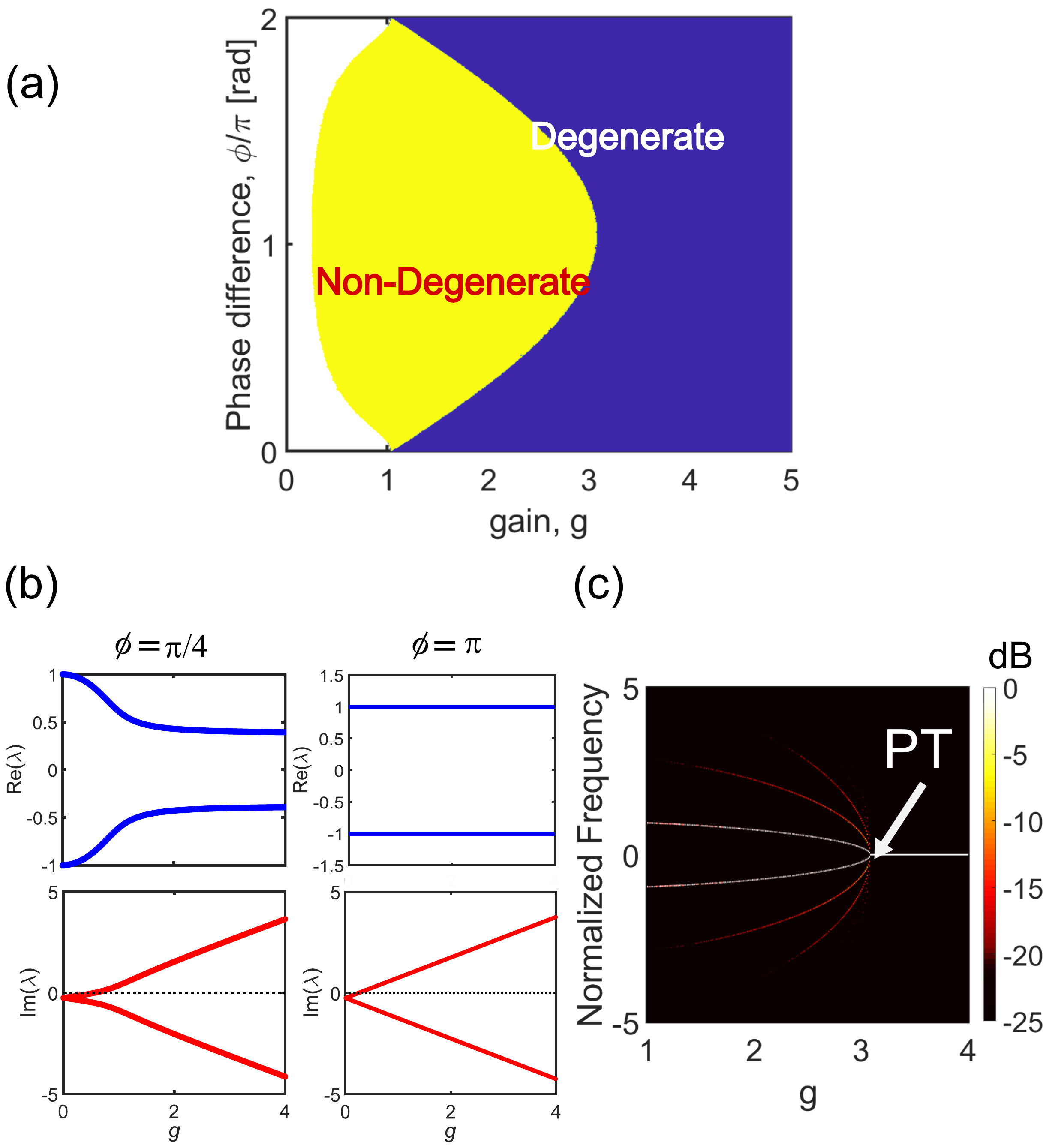}
\caption{\label{fig: schematic} \textbf{Occurrence of nonlinearity induced phase transition}. a) Phase diagram of coupled OPO driven with pump of similar strength $g$ with a relative phase difference $\phi$ . The phase diagram clearly shows the presence of two phases of oscillation above threshold, namely the non-degenerate and degenerate. The white region indicates that the coupled OPO system is below threshold. The phase diagram is obtained by solving the coupled nonlinear equations for the OPOs (Eq 1) including the gain saturation. b) Linearized analysis predicts the possibility of non-degenerate oscillation only, however a phase transition into degenerate phase can be engendered when accounting for the back-conversion nonlinearity. Two representative cases for $\phi=\pi/4$ and $\phi=\pi$ are shown. c) Nonlinearity induced phase transition from non-degenerate to degenerate (for $\phi=\pi$) highlighting the soft/super-critical nature of phase-transition. The colorbar represents the spectral intensity in dB scale.}
\end{figure}
The phase sensitive nature of parametric gain provides an additional tuning knob in the form of phase difference between the two driving pumps ($\phi$) that do not exist in the conventional phase insensitive gain/loss based non-Hermitian systems. Figure 3a illustrates the solution space as the phase difference is varied identifying the degenerate and the non-degenerate oscillation regimes. \
\begin{figure}[!h]
\centering
\includegraphics[width=0.5\textwidth]{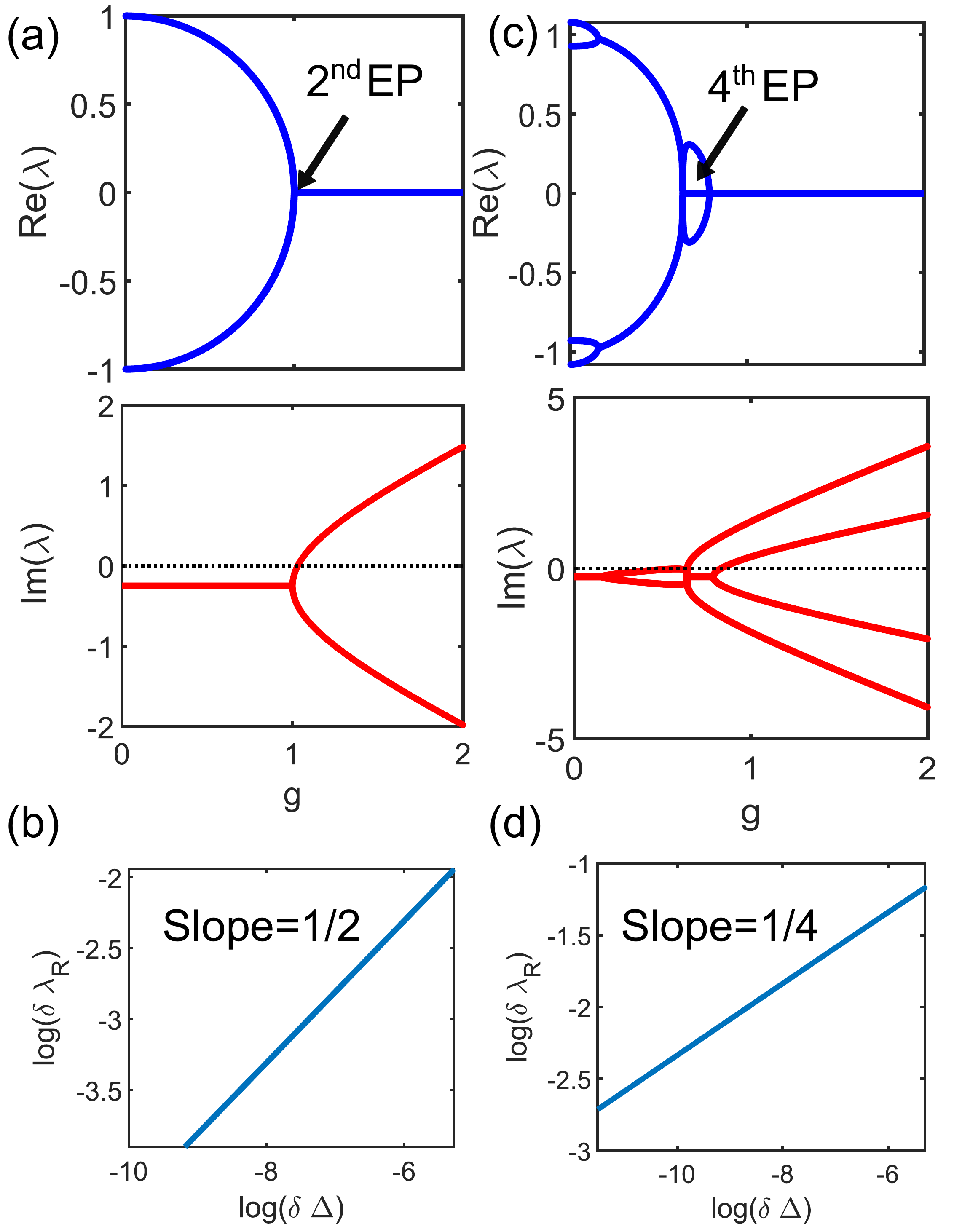}
\caption{\label{fig: schematic} \textbf{Different orders of EP realized in coupled OPO}. a) Second order EP. Parameters used are $f=g$, $\kappa =1$, and $\gamma =0.25$.  b) Dependence of the spectral splitting $(\delta Re(\lambda))$in the vicinity of EP, when a perturbation in the form of detuning $(\Delta_{1}=\delta \Delta)$is applied. It varies as $\delta Re(\lambda) = (\delta \Delta)^{\frac{1}{2}}$.  c) Fourth order EP. Parameters used are $f=2g$, $\kappa =1$, $\Delta_{1}=0.1501$ and $\gamma =0.25$.  d) Dependence of the spectral splitting $(\delta Re(\lambda))$ in the vicinity of EP, which varies as $\delta Re(\lambda) = (\delta \Delta)^{\frac{1}{4}}$.}
\end{figure}

The order of exceptional point determines the rate of eigenvalues splitting in the presence of a perturbation away from EP \cite{miri2019exceptional}. If the perturbation appears in the form of detuning ($\delta \Delta$), then the splitting depends as: $\delta (Re(\lambda))= (\delta \Delta)^{\frac{1}{n}}$, where $n$ is the order of EP. This leads to enhanced sensitivity in the proximity of an EP, which is given by $\frac{d (\delta (Re(\lambda)))}{d (\delta \Delta)} \sim (\delta \Delta)^{\frac{1-n}{n}}$. This sensitivity function diverges at EP, which is the basis for enhanced sensitivity of EP based sensors \cite{hodaei2017enhanced, lai2019observation, chen2017exceptional}. This scaling law arising due to the branch point singularity nature of non-Hermitian degeneracies does not arise in case of Hermitian degeneracies characterized by the diabolical points. We present the occurrence of both a second order EP and higher order EP (4 th order) in the coupled OPO system. Second order EP is accompanied by the collapse of eigenvalues and eigenvectors in pairs and is shown in Fig. 4a and 4b by considering $f=g, \Delta_{1}=\Delta_{2}=0$. We identify a family of higher order exceptional points \cite{sup}, by biasing the coupled OPO's at suitable detuning. In Fig. 4c and 4d we considered $f=mg, \Delta_{2}=0$, where $m$ is the parameter describing the family of exceptional points which determines the critical $g$ and $\Delta_{1}$ for the occurrence of the 4th order EP. In this case, four eigenvectors and eigenvalues coalesce resulting in higher order dependence of sensitivity.  This enhanced sensitivity of the 4th order EP is reflected in the slope of the log-log plot in Fig. 4d as compared to the case in Fig. 4b corresponding to a second order EP. \\

The instantaneous nature of parametric gain and the ability to modulate the gain by applying phase/ amplitude modulation to the pump opens unprecedented avenues in exploring time modulated dynamic non-Hermitian systems in the coupled OPO arrangement. Time periodic Floquet non-Hermitian systems have been utilized to tailor the EP and realize re-configurable non-Hermitian systems with an enriched phase space depending on the amplitude and frequency of the modulation \cite{luo2013pseudo,li2019observation, chitsazi2017experimental}. Previous demonstrations relied on periodically modulating the coupling to realize Floquet driven systems. Parametric non-Hermitian systems enable us to modulate the gain instead of the coupling, by modulating the pump and realize tunable Floquet EP. In Fig. 5a and 5b we explore Floquet control of EP when the pump is amplitude modulated as: $g=g_{0}+F\textrm{sin}(\omega t)$. The eigenvalues of the Floquet periodic system can be extracted by analyzing the associated Monodromy matrix. As shown in Fig. 5b with increasing values of the pump amplitude modulation parameter $F$, the EP is progressively shifted to higher values of $g_{0}$.\\

Similarly, we can dynamically encircle the EP by periodically modulating the parametric gain. Dynamical encirclement involves adiabatically tracing a close path in the parameter space enclosing an EP, which has been utilized to realize robust and asymmetric switching \cite{doppler2016dynamically}, non-reciprocal energy transfer \cite{xu2016topological} and omni-polarizer \cite{hassan2017dynamically}.  However these promising results have only been demonstrated in lossy systems \cite{doppler2016dynamically,yoon2018time}, due to the stringent requirement of non-Hermitian system tunability. Here we propose that the tunable nature of the parametric gain provides a very promising platform to realize these chiral dynamics that is contingent to the topological structure of the EP. We perform adiabatic encirclement in the parametric space (Fig. 5c and 5d) of detuning and gain by undergoing the following adiabatic evolution: $f=g=g_{0}+r\textrm{cos}(\omega t)$ and $\Delta_{1}=r\textrm{sin}(\omega t)$, where $r$ is the radius of encirclement, and $g_{0}=\kappa$ is the EP. Due to the breakdown of adiabaticity in non-Hermitian system, we obtain an asymmetric/ chiral behavior, where the final state at the end of the encirclement, depends on the direction of the loop and is independent of the starting point. The distinct outcome by parametrically traversing a loop enclosing the EP counter-clockwise (Fig. 5c) and clockwise (Fig. 5d) is shown. \\ \\
\begin{figure}[!h]
\centering
\includegraphics[width=0.5\textwidth]{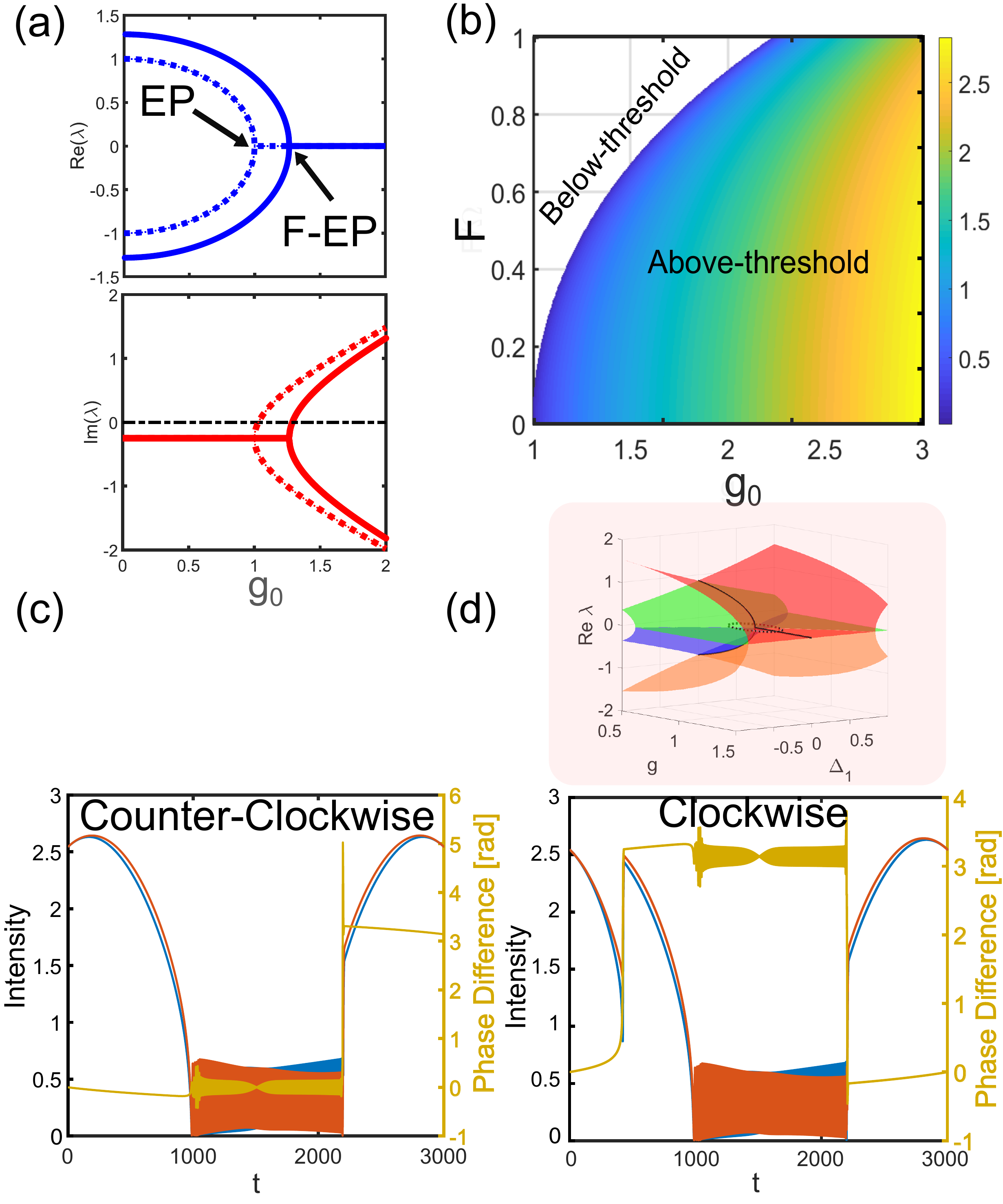}
\caption{\label{fig: schematic}\textbf{Periodically modulated coupled OPO}. a) Floquet control of EP. Real and Imaginary parts (solid lines) of the Floquet exponent of the associated monodromy matrix. Floquet EP (F-EP) can be tuned from the static EP  (dotted lines represents the eigenvalues for the static system without periodic modulation). Parameters for the pump amplitude modulation used are $F=5, \omega =10$.    b) Tuning the F-EP by varying $F$, the amplitude modulation depth. The threshold in the coupled OPO system can be varied by changing $F$ (white region corresponds to below threshold). The color-bar represents gain $\left (\textrm{Im}(\lambda)+\gamma \right)$. c) Adiabatically encircling the EP, and the emergence of chirality. Counter-clockwise encirclement, and the system ends up in a different final state. d) Clockwise encirclement where an abrupt jump occurs during the evolution, and the system returns to the initial state at the end of the encirclement. Highlighted is shown the Riemann eigen-surfaces. The dotted loop represents the encirclement trajectory on the $\Delta_{1} - g$ parameter space. The black solid line indicates the eigen-frequency splitting for $\Delta_{1}=0$. Parameters used are: $r=0.2, \omega = \frac{2\pi}{3000}$.}
\end{figure}
OPOs have been the workhorse for generating quantum states of light for decades \cite{wu1986generation}, and coupled OPOs have also been predicted to exhibit nonclassical properties  \cite{olsen2005entanglement}. When we approach the EP from below threshold the vacuum fluctuations in the quadratures of the intra-cavity field can be squeezed below the standard noise limit. We assume the vacuum fluctuations entering the cavity from different open channels, to be delta correlated white Gaussian noise and obtain the power spectral density of the output quadrature fields via a linearized analysis of the Langevin equations  \cite{chembo2016quantum}. The formalism is derived in \cite{sup}. Figures 6a and 6b show that there exists a bandwidth where the intra-cavity field is squeezed as we approach the EP. The reduced noise in one quadrature is accompanied by increased uncertainty (anti-squeezing) in the conjugate quadrature. Although,  the maximum squeezing attainable in the vicinity of EP is 3dB \cite{sup}, it can potentially allow to combine the high sensitivity of EP and the reduced uncertainty in parametric EP, to realize unparalleled sensing capabilities in an optimum sensing arrangement. More so, one can tune the squeezing response by changing $\kappa$ in coupled OPO as shown in Fig. 6b, thereby operating at a frequency where external/technical noise of the sensing system is minimum. In response to a perturbation in the form of detuning, only the optimum quadrature for squeezing is rotated, still preserving the noise reduction property \cite{sup}. In this regard our parametric EP can pave the way for ultra-sensitive detection with high SNR in shot noise limited detection scenarios.     
\begin{figure}[!h]
\centering
\includegraphics[width=0.5\textwidth]{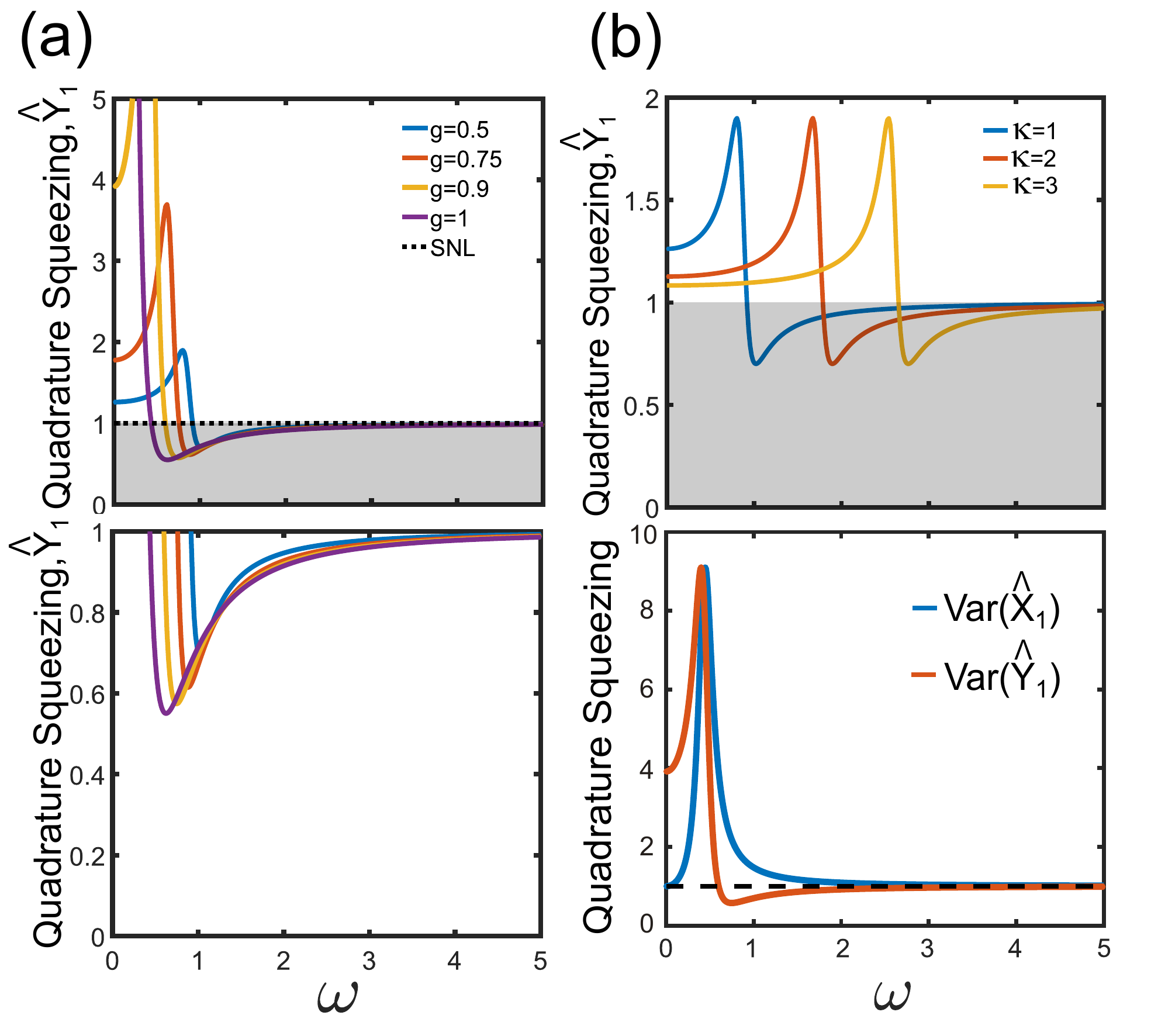}
\caption{\label{fig: schematic} \textbf{Squeezing near the parametric EP in coupled OPO}. a) Quadrature squeezing spectrum  as we approach the EP by varying $g$. There exists a bandwidth where the squeezing is below the standard noise limit indicated by the black shaded region and bounded by the black dotted line. A zoomed in version of the same is shown right below. b) The squeezing spectrum can be tuned by varying the coupling strength $\kappa$ in the coupled OPO system. $g$ is kept equal to 0.5$\kappa$.  Below is shown the squeezing and anti-squeezing in the conjugate quadratures for $\kappa =1, g=0.9, \gamma =0.1, \rho =0.9$. }
\end{figure}

In summary, we have presented EP in parametrically driven coupled OPOs. We identified the presence of two distinct phases of oscillation, namely the degenerate and the non-degenerate, and have shown nonlinearity induced phase transition. We discussed the potential benefits of tunable parametric gain in realizing dynamically modulated non-Hermitian systems. Non-classical behaviour of the parametric EP is presented and its implications in highly sensitive/ high SNR detection is highlighted. \\

Recent developments in realization of large-scale time-multiplexed OPO networks \cite{marandi2014network} and integrated lithium niobate based devices \cite{wang2018integrated} can be ideal candidates for experimental realization of the presented concept. Entanglement can be used as a resource for increasing the sensor performance \cite{degen2017quantum} based on parametric EP.  An optimum sensing arrangement guided by Quantum Fischer information calculations needs to be designed in order to obtain high SNR sensing from parametric EP \cite{lau2018fundamental, zhang2019quantum}. Also, the enhancement provided by higher order parametric EP and the limits of sensors based on them including their dynamic range is worth exploring and will be subjects for future investigations.   
\begin{acknowledgments}
The authors gratefully acknowledge support from ARO Grant No. W911NF-18-1-0285 and NSF Grant No. 1846273 and 1918549. S. F. acknowledges the support of a Vannevar Bush Faculty Fellowship from the U. S. Department of Defense (Grant No. N00014-17-1-3030). The authors wish to thank NTT Research for their financial and technical support.
\end{acknowledgments}

\nocite{*}
\bibliography{cr}

\end{document}